\documentclass[12pt]{iopart}

\usepackage{amssymb}
\usepackage{graphics,graphicx}
\usepackage{color}
\newcommand{\beq}{\begin{equation}}
\newcommand{\eeq}{\end{equation}}

\newcommand{\bea}{\begin{eqnarray}}
\newcommand{\eea}{\end{eqnarray}}

\newcommand{\ba}[1]{\begin{array}{#1}}
\newcommand{\ea}{\end{array}}
\newcommand{\ben}{\begin{enumerate}}
\newcommand{\een}{\end{enumerate}}
\newcommand{\bit}{\begin{itemize}}
\newcommand{\eit}{\end{itemize}}
\newcommand{\bde}{\begin{description}}
\newcommand{\ede}{\end{description}}

\renewcommand{\[}{\left[}
\renewcommand{\]}{\right]}

\begin{document}

\title{Layer-layer competition in multiplex complex networks}

\author{Jes\'us G\'omez-Garde\~nes$^{1,2}$, Manlio De Domenico$^3$, Gerardo Guti\'errez$^4$, Alex Arenas$^3$ and Sergio G\'omez$^3$}

\address{$^1$ Departamento de F\'{\i}sica de la Materia Condensada, Universidad de Zaragoza, 50009 Zaragoza (Spain)}
\address{$^2$ Instituto de Biocomputaci\'on y F\'{\i}sica de Sistemas Complejos, Universidad de Zaragoza, 50018 Zaragoza (Spain)}
\address{$^3$ Departament d'Enginyeria Inform\`atica i Matem\`atiques, Universitat Rovira i Virgili, 43007 Tarragona (Spain)}
\address{$^4$ Departamento de Ciencias Aplicadas, Instituto Tecnol\'ogico Metropolitano, 354 Medell\'{\i}n (Colombia)}

\begin{abstract}
The coexistence of multiple types of interactions within social, technological and biological networks has moved the focus of the physics of complex systems towards a multiplex description of the interactions between their constituents. This novel approach has unveiled that the multiplex nature of complex systems has strong influence in the emergence of collective states and their critical properties. Here we address an important issue that is intrinsic to the coexistence of multiple means of interactions within a network: their competition. To this aim, we study a two-layer multiplex in which the activity of users can be localized in each of the layer or shared between them, favoring that neighboring nodes within a layer focus their activity on the same layer. This framework mimics the coexistence and competition of multiple communication channels, in a way that the prevalence of a particular communication platform emerges as a result of the localization of users activity in one single interaction layer. Our results indicate that there is a transition from localization (use of a preferred layer) to delocalization (combined usage of both layers) and that the prevalence of a particular layer (in the localized state) depends on their structural properties.
\end{abstract}

\maketitle

\section{Introduction}
During the last 15 years many statistical physics methods and nonlinear models have suffered a reformulation in order to take into account non-regular interaction patterns \cite{rev1,rev2,rev3}. This reformulation is rooted in the availability of datasets capturing the relationships among the constituents of macroscopic systems of diverse nature, such as technological, biological and social ones, and their successful description in terms of complex networks \cite{book1,book2}. As a result, many important collective phenomena taking place in these systems, such as synchronization \cite{sync}, epidemics \cite{epic}, cooperation and consensus \cite{social} among others, have been revisited under the paradigm of complex networks \cite{rev4,rev5}.

As data gathering techniques increase their resolution new properties of the interaction patterns in complex systems are captured. Main features include the spatial, temporal and multiplex nature of interaction networks. This latter ingredient has greatly focused the attention of network science in the last years leading to a number of works about the structure and dynamics of multilayer and multiplex networks \cite{jcn,physrep}.

\begin{figure}[!t]
  \centering\includegraphics[width=4.3in,clip=true]{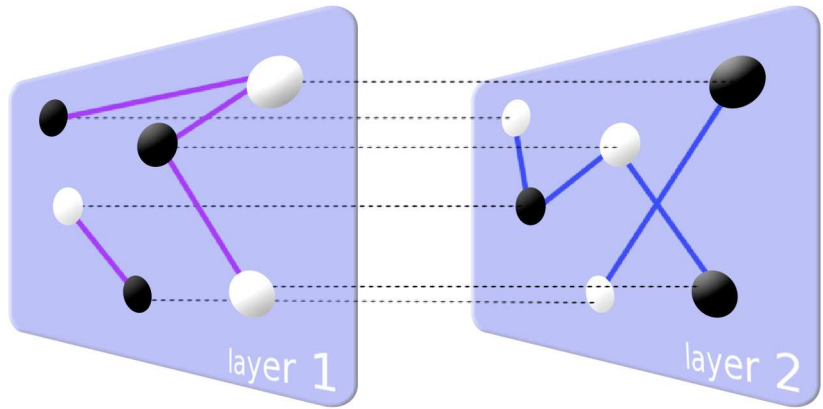}
  \caption{Schematic representation of a two-layers multiplex network. The multiplex is composed of $6$~nodes so that each node appears in each layer and it is connected (dashed lines) with its representation in the other layer. When layers compete, each user must choose the layer in which is active (black), thus remaining inactive (white) in the other one.}
  \label{fig1}
\end{figure}

Multiplex networks \cite{prx} are often described as the framework for capturing the interactions among a set of elements (nodes) when these interactions can take different forms or be established through different means, each of them defining a network often referred to as interaction layer. Thus, a multiplex network can be seen as a collection of layers so that each node is represented in all of them (see. Fig.~\ref{fig1}). Typical examples of multiplex networks are transportation systems \cite{cardillo,navigability,barthelemy,cardillo2}, in which different transportation modes can be used to travel between cities or urban areas, or social systems \cite{szell}, in which individuals can choose among different means and communication platforms for interacting with each other.

Most of the times the different interaction layers forming the multiplex coexist in a competitive way \cite{kolja1,kolja2,coreanos}. For instance think of two of the most important applications for mobile communication, such as {\em WhatsApp} and {\em Line}, here the competition relies on the usage of each platforms. The more users decide to use one platform the more value has the platform. From the point of view of users the final choice between these two platforms relies on two main issues. Obviously, the intrinsic quality of the platform plays a key role in the final decision of individuals. However, there is social added value that comes from the degree of usage  of the platform among the acquaintances of an individual. It is thus interesting how this local and context-driven decisions affect the onset of a collective state, here represented as the localization of the multiplex activity in one of its layers.

The article is organized as follows. In Sec.~\ref{math} we present the mathematical formulation of the multiplex network model and derive the relevant equations to be solved. In Sec.~\ref{num} we show the numerical results corresponding to the competition between different  interaction layers. Finally, in Sec.~\ref{conc} we round off the manuscript with the conclusions of our work and the future perspectives that it may open.

\section{Mathematical formulation of multiplex layers competition}
\label{math}

The physical framework used to study the layer-layer competition relies on a two-states model (reminiscent of an Ising-like model but much simpler). In this case, nodes are two-states (up and down) systems and the interaction neighborhood of a node depends on its (up or down) state. Thus, at the macroscopic scale, this translates into the existence of two interaction layers, one associated with the up state and another with the down one. In the following we first describe in~\ref{sec:general} the general Hamiltonian capturing the interactions between the nodes of a multiplex of $L$~layers to particularize in~\ref{sec:2-layers} to the case of competitive interactions in a two-layers multiplex.


\subsection{General formulation: Multiplexes of $L$~layers}
\label{sec:general}
In general, a multiplex network is composed of $L$~network layers of $N$~nodes each. Since each individual~$i$ is represented in each of the~$L$ networks, each pair of networks~$\alpha$ and~$\beta$ are interconnected by $N$ links connecting the pair of nodes that represent the same individual $i$. This setup can be seen as a collection of independent layer or platforms available for the communication between individuals (such as  WhatsApp, Line, Tango, etc) or for data-sharing (such as Dropbox, iCloud, Box, etc), being the nodes of the layers, the users, and the links within a layer the connections established by the users via a particular platform.

The availability of different platforms oriented towards the same goal poses a natural competition for the choice of the users. The essential ingredients of this competition can be casted in a mathematical formulation in which the state of a node $i$ in layer $\alpha$ can be explained as the probability that node $i$ is active in layer $\alpha$ (for communication platforms) or the fraction of resources that node $i$ share with its neighbors in this layer (for data-allocation systems). In this way, the state vector of a node $i$ in the interconnected multilayer network is denoted by $\vec{p}_{i}\equiv(p_{i}^{(1)},\ldots,p_{i}^{(L)})$, together with the constraints that the sum of the probabilities of finding a node active in each layer is equal to $1$:
\begin{equation}
  \label{constraint}
  \sum_{\alpha=1}^{L}p_{i}^{(\alpha)}=1,\quad i=1,2,\ldots,N.
\end{equation}

The state of the multiplex can be represented by a $L\times N$ matrix $\mathbf{P}\equiv \[(\vec{p}_{1})^{\dag},\ldots,(\vec{p}_{N})^{\dag}\]$. The \emph{state matrix}, accounting for the intensity of interaction between the nodes is given by all possible products between the states, {\em i.e.}, by the $LN\times LN$ matrix $\mathbf{\Sigma} = \mathbf{P}\otimes \mathbf{P}^{\dag}$ with elements $\Sigma^{(\alpha\beta)}_{ij}=p_i^{(\alpha)}p_j^{(\beta)}$.
In addition, we define ${\mathbf J}$ as a $LN\times LN$ \emph{interaction matrix} capturing both intra-layer and inter-layer links:
\begin{equation}
  \mathbf{J}=\bigoplus \mathbf{W}^{(\alpha)}+\mathbf{D}\otimes\mathbf{I}\;,
\end{equation}
where $\mathbf{I}$ indicates the $N\times N$ identity matrix, $\mathbf{D}$ is the $L\times L$ matrix accounting for the network of layers \cite{diffsole} and $W^{(\alpha)}_{ij}$ is the weight of the interaction between nodes $i$ and $j$ in layer $\alpha$. More specifically, this topology describes nodes that are present in multiple layers simultaneously and inter-layer connections are allowed only between a node and its counterparts in the other layers. If $\alpha$ and $\beta$ are indices indicating two given layers, the block matrix structure of $\mathbf{J}$ can be indexed by four indices, two for nodes and two for layers, {\em i.e.}, by $J^{(\alpha\beta)}_{ij}$.

Therefore, given the ensemble of all possible states, {\em i.e.}, the set $\{\mathbf{P}\}$ of all matrices satisfying constraints Eq.~(\ref{constraint}), we can define the Hamiltonian of a specific configuration $\mathbf{P}$ as:
\begin{equation}
  \label{genhamiltonian}
  \mathcal{H}(\mathbf{P})=-\sum_{\alpha,\beta=1}^{L}\sum_{i,j=1}^{N}J^{(\alpha\beta)}_{ij}p_i^{(\alpha)}p_j^{(\beta)}\;.
\end{equation}


\subsection{Competition in $2$-layers multiplexes}
\label{sec:2-layers}

For the sake of simplicity, hereafter we consider the case of the competition in a multiplex composed of two layers, i.e., $L=2$. In this specific case, following Eq.~(\ref{constraint}), the state of a node is completely determined by its probability of being active in one of the two layers, {\em e.g.}, the first one. Thus, the state of the whole multiplex can be described by the vector $\vec{p}\equiv(p_1,\ldots,p_N)$, where for simplicity we have omitted  the layer index explicitly. Moreover, we also consider a uniform and undirected connection between the two layers, so that the interaction strength of a node $i$ in layer $1$ and its counterpart in layer $2$ is captured by the parameter $J_{x}$.

To incorporate the competition between layers we consider two essential ingredients of the interactions at the local level. On one hand, the communication between two agents that are connected within one of the layers is more efficient when both of them are always active in this layer or when they allocate all of their shared resources in the same platform. However, since the sets of contacts an individual has in the two layers are, in principle, different, by splitting the activity between the two layers an individual will increase the number of simultaneous contacts. The Hamiltonian capturing these two ingredients can be obtained from the general one in Eq.~(\ref{genhamiltonian}) as:
\begin{equation}
  \label{hamiltonian2}
  H(\vec{p})=-\sum_{i,j=1}^{N}W^{(1)}_{ij}p_{i} p_{j}-\sum_{i,j=1}^{N} W^{(2)}_{ij}(1-p_{i})(1-p_{j})-2J_x\sum_{i=1}^{N} p_{i}(1-p_{i}).
\end{equation}
From this Hamiltonian it becomes clear that the first two terms in the right are those favoring the localization of the activity of each individual in layer one and two respectively. In its turn, the third term favors the splitting of the activity of each individual.

The relative importance of this third term with respect to those favoring the localization of the activity within a single layer is controlled by the inter-layer coupling $J_x$. Note that the limit $J_x\gg 1$ means that nodes are prone to combine their activity in both layers which, for instance, in the case of data sharing or mobile communication platforms would represent information (pictures, files, tweets, etc) that can be easily transferred from one platform to the other one. On the contrary, the case $J_x\ll 1$ implies that a simultaneous use of platforms is hard to achieve.


\section{Results}
\label{num}

Having introduced the mathematical framework, our goal is to study the competition between the two layers as a function of the inter-layer strength $J_x$ and the structural patterns of each of the network layers. To this aim, it is useful to check the behavior in the two asymptotic limits: $J_x\gg 1$ and $J_x=0$. First, when $J_x\gg 1$ the first two terms in the Hamiltonian Eq.~(\ref{hamiltonian2}) become negligible, so that the configuration of minimum energy is achieved for $p_i=1/2$ $\forall i$, {\em i.e.}, when the individuals split their activity between the two layers. On the other hand, for $J_x=0$ the multiplex becomes a set of two independent networks and the configurations localized in the first layer ($\vec{p}=\vec{1}$, {\it i.e.} $p_i=1$ $\forall i$) and the second one ($\vec{p}=\vec{0}$) compete. In this case the minimum energy configuration is achieved by concentrating all the activity in the layer $\alpha$ with the largest total strength:
\begin{equation}
  s^{(\alpha)}=\sum_{i,j=1}^N W_{ij}^{(\alpha)}\;.
\end{equation}
For the particular case of unweighted networks this means that the layer with the largest average degree (or largest number of links) will concentrate all the activity when $J_x=0$. From now on, we will consider that $s^{(1)}>s^{(2)}$ so that in the absence of inter-layer interactions the activity focuses on the first layer: $\vec{p}=\vec{1}$.

Considering these two asymptotic behaviors we are thus interested in characterizing the transition from the localized activity regime at small values of inter-layer coupling $J_x$ to that of mixed one for large $J_x$. A first proxy is to check when the fully localized solution ({\em e.g.} $\vec{p}=\vec{1}$) ceases to be the one with the minimum energy $H=-s^{(\alpha)}$ ($\alpha=1$ in the case of $\vec{p}=\vec{1}$). To this aim, we calculate the gradient of the multivariate Hamiltonian $H(\vec{p})$:
\begin{equation}
  \frac{\partial H}{\partial p_i}=-2\sum_{j=1}^N W^{(1)}_{ij}p_j+2\sum_{j=1}^N W^{(2)}_{ij}(1-p_j)-2J_x(1-2p_i)\;.
\end{equation}
The values of these derivatives for the localized solution in the first layer ($\vec{p}=\vec{1}$) become:
\begin{equation}
  \left.\frac{\partial H}{\partial p_i}\right|_{\vec{p}=\vec{1}}=2\left(J_x-s_i^{(1)}\right)\;,
  \label{derivsign}
\end{equation}
where $s_i^{(\alpha)}=\sum\limits_{j=1}^N W^{(\alpha)}_{ij}$ is the strength of node $i$ in layer $\alpha$. For values of $J_x$ smaller than the strength $s_i^{(1)}$ of all nodes in the first layer, the derivatives at $\vec{p}=\vec{1}$ are all negative and thus the gradient points to the interior of the hypercube $\[0,1\]^N$, which contains all the possible feasible states of the system. This means that the energy of the system around $\vec{p}=\vec{1}$ is always increased for any small change of $\vec{p}$ inside the hypercube, showing that $\vec{p}=\vec{1}$ has minimum energy whenever the inter-layer coupling $J_x$ is below its critical value:
\begin{equation}
  J_x^c=\min_{i=1,\ldots,N}(s_i^{(1)}) = s_{\min}^{(1)}\;.
\end{equation}
For unweighted networks it reduces to the minimum degree of the nodes in the first layer, $k_{\min}^{(1)}$. Above this critical inter-layer coupling $J_x^c$ the minimum energy moves from $\vec{p}=\vec{1}$ to a new position inside the hypercube $\[0,1\]^N$, thus starting to distribute the activity between the two layers.

From a mathematical point of view, the finding of the state with minimal energy is a quadratic (the Hamiltonian) programming problem with linear equality (normalization of the probabilities) and inequality (probabilities in range $[0,1]$) constraints. In general, for two layers, the candidate minimum of the Hamiltonian is calculated by setting $\partial H/\partial p_i=0$ $\forall i$, which can be expressed as the following linear system:
\begin{equation}
  \label{minimum}
  \left[2J_x{\bf I}-\left({\bf W}^{(1)}+{\bf W}^{(2)}\right)\right]\vec{p}=J_x\vec{1}-\vec{s}^{(2)}\,.
\end{equation}
However, its solution $\vec{p}^{\star}$ does not always fulfill the constraints, does not constitute a minimum, or even both conditions fail at the same time. When any of these happens, the minimum is placed in the boundaries of the $\[0,1\]^N$ hypercube, with at least one probability equal to~1 or~0. Supposing $\vec{p}^{\star}$ is inside the hypercube, it is a minimum if the Hessian matrix, with components
\begin{equation}
  \label{hessian}
  \frac{\partial^2 H}{\partial p_i\partial p_j}=2\left(2J_x\delta_{ij}-W_{ij}^{(1)}-W_{ij}^{(2)}\right)\;,
\end{equation}
is positive definite. When the Hessian is not positive definite, the quadratic programming problem becomes NP-hard, and no polynomial time algorithm is known to solve it. Since the Hessian is proportional to the matrix of the system in Eq.~(\ref{minimum}), a positive definite Hessian implies the system has a non-singular matrix and thus a unique solution.

Summarizing, we proceed as follows to find the state with minimum energy. For each value of the coupling $J_x$, we first solve Eq.~(\ref{minimum}) and obtain a solution $\vec{p}^{\star}$. Then, we check if $\vec{p}^{\star}\in \[0,1\]^N$ and if all the eigenvalues of the Hessian Eq.~(\ref{hessian}) are positive. If both conditions are fulfilled, $\vec{p}^{\star}$ is the ground state for this value of the coupling and we have finished. Otherwise, a heuristics is needed to solve the problem. We have chosen Particle Swarm Optimization \cite{pso} for its simplicity, ability to cope with continuous and bounded variables, and outstanding performance in many fields \cite{psoj}.

In order to represent the ground state of the Hamiltonian for each value of $J_x$ we make use of magnetization $M(\vec{p})$:
\begin{equation}
  \label{magnetization}
  M(\vec{p})=\frac{1}{N} \sum_{i=1}^N (2 p_i - 1)\;,
\end{equation}
that characterizes the level of activity between layers: $M=1$ when all the activity is concentrated in the first layer ($\vec{p}=\vec{1}$) and $M=-1$ when it is concentrated in the second layer ($\vec{p}=\vec{0}$). Obviously, when activity is shared between the two layers ($\vec{p}=\vec{0.5}$) the magnetization vanishes, $M=0$.

\begin{figure}[!tb]
  \centering\includegraphics[width=12cm,clip=true]{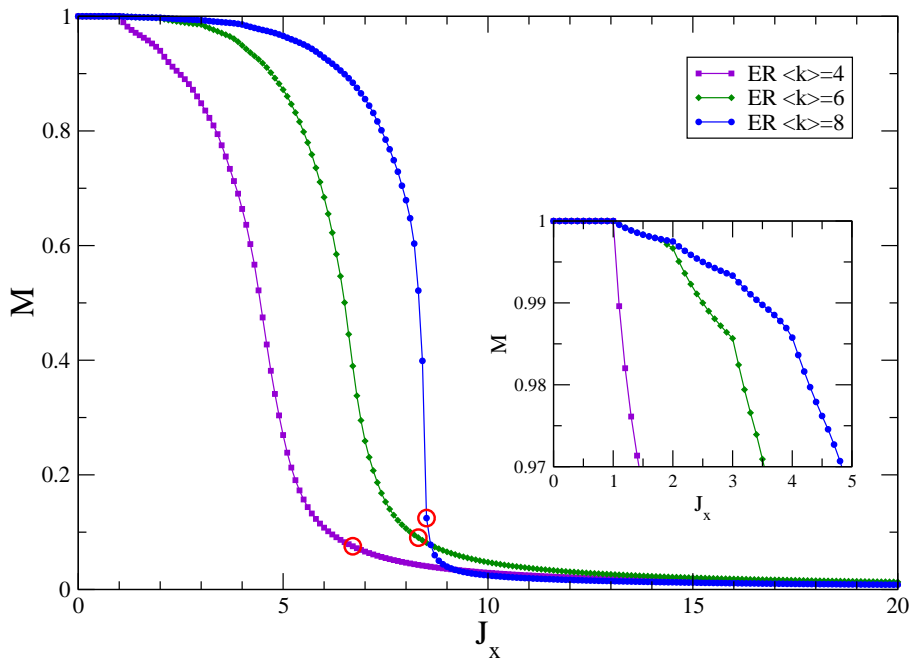}
  \caption{Magnetization for multiplex networks consisting of two Erd\H{o}s-R\'{e}nyi (ER) layers and varying mean degree. All ER networks have 200~nodes and minimum degree $k_{\min}=1$. The red circles indicate the minimum values of $J_x$ so that minimum energy configurations can be calculated from Eq.~(\ref{minimum}). For values of $J_x$ to the left of the red circle, PSO has been used to compute the solutions. In the inset, zoom to show when activity starts to be distributed in both layers.}
  \label{fig2}
\end{figure}

In Fig.~\ref{fig2} we show the magnetization as a function of $J_x$ for three multiplex networks composed of two Erd\"os-R\'enyi (ER) layers. The layers of the three multiplex networks have average degree $\langle k\rangle=4,$ $6$ and $8$ respectively. The layers were produced by means of the algorithm introduced in \cite{ER} that allows to interpolate between ER and scale-free networks and, as a byproduct of the procedure, enables to control the minimum degree of the resulting graphs. In this way we have set the minimum degree to $k_{\min}=1$ in the three cases. However, the method in \cite{ER} produces (for a given value of $k_{\min}$) networks of identical strength. Thus, we take the first network layer an add $0.05\times N$ links at random to ensure that $s^{(1)}>s^{(2)}$.
The three curves in Fig.~\ref{fig2} display the transition from localized activity for small $J_x$ values to mixed one for large ones. Interestingly, the inset shows that localized activity is lost as soon as $J_x=1=k_{\min}$ in agreement with our former estimation. From this point the decay of $M$ is slower for those multiplexes with larger mean degree $\langle k\rangle$ in the layers.

\begin{figure}[!tb]
  \centering\mbox{\includegraphics[width=12cm,clip=true]{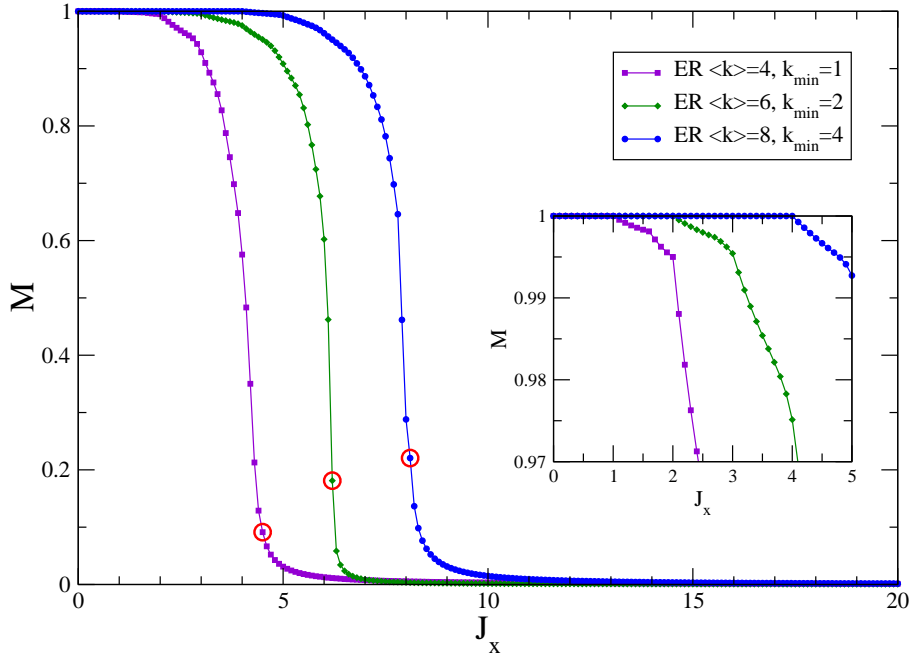}}
  \caption{Magnetization for multiplex networks consisting of two Erd\H{o}s-R\'{e}nyi (ER) layers and varying mean and minimum degrees. All ER networks have 200~nodes. The red circles indicate the minimum values of $J_x$ so that minimum energy configurations can be calculated from Eq.~(\ref{minimum}). For values of $J_x$ to the left of the red circle, PSO has been used to compute the solutions. In the inset, zoom to show when activity starts to be distributed in both layers.}
  \label{fig3}
\end{figure}

These results are further corroborated in Fig.~\ref{fig3}. In this case we show again three two-layers multiplexes (again composed of coupled ER-ER networks) with the same average degrees as in Fig.~\ref{fig2} ($\langle k\rangle=4$, $6$ and $8$) but with different values of the minimum degree, $k_{\min}=1$, $2$ and $4$ respectively. This latter feature is revealed in the inset of the plot where we show that the state of localized activity ($M=1$) is no longer the ground state of the multiplex as soon as $J_x>k_{\min}=1,$ $2$ and $4$. On the other hand, as in Fig.~\ref{fig2}, the final mixed activity state is achieved first for those multiplexes with smaller average degree.

In both Figs.~\ref{fig2} and~\ref{fig3} the insets show that the evolution of $M(J_x)$ close to $M=1$ shows different cusps. In particular, these cusps appear at integer values of $J_x$, which at the same time correspond to the values at which the number of nodes for which Eq.~(\ref{derivsign}) changes sign increases. The effect is similar to that at $k_{\min}$: many nodes have $p_i = 1$ until $J_x$ is equal to $k_i$ in the first layer. This rule is exact only for nodes with $k_i = k_{\min}$, but also holds for an important fraction of nodes when $k_i > k_{\min}$. The collective effect is reflected in the form of cusps in the curve $M(J_x)$.


\section{Conclusion}
\label{conc}

In this work we have introduced a model to analyze how layers compete for the activity of users in a multiplex network inspired in the simultaneous interplay of communication and data-sharing online platforms. To this aim, we have focused on a multiplex composed of two layers and we have relied on a two-states model in which each of the two states of a node are associated to be active in the top layer and the bottom one respectively. At variance with the usual Ising model in a network \cite{Ising}, here a node interacts only with those neighbors in the layer it is active. We have set two competing mechanisms, one favoring activity localization and another favoring the splitting of the node's activity between the two layers. On one hand, if a node and all its neighbors in a layer are active in it, this would favor an efficient communication between them. However, when a node splits its activity between the two layers this would favor the passage of information from a neighbor in one layer to a neighbor in the other one, thus maximizing the outreach of information. The competition between these two mechanisms is controlled by the inter-layer coupling $J_x$ which can be seen as the ability that node has to pass information from one layer to the the other.

Our results show that, regardless of the average connectivity and total strength of the leading layer (the one that focuses all the activity for small inter-layer coupling) it is its minimum degree what causes that nodes start to use the other layer, {\em i.e.} controls the onset of the transition from localized to mixed activity. On the other hand, it is the average degree of the leading layer what controls when the state of full mixing is reached. These two results point out that the transition from localized to mixed activity occurs via a cascade from poorly connected nodes in the leading layers (the ones that obtain more benefits from leaving first the leading layer) to those highly connected ones (being the ones that are less prone to leave the layer in which they are well-connected). Thus, the larger the average degree of the leading layer, the more inter-layer coupling is needed to persuade all the degree classes to leave the localized state.

We expect that the simple model introduced here will stimulate more research about the coexistence and competition of interaction layers in multiplex networks making possible the characterization of how and when the coexistence of different layers in real multiplex systems is possible. Future research avenues include the study of other types of layer topologies and the presence of correlations between the degrees of a node in different layers. Moreover, a more challenging problem is the competition in multiplexes composed of more than two layers, characterized by the Hamiltonian in Eq.~(\ref{genhamiltonian}). It is clear, that the existence of multiple parameters for the interaction between the $L$ layers poses a mathematical and computational difficulty. On the other hand, this general framework provides with the interesting scenario in which many different transitions between the localization in different layers are observed due to the multiple competition between them. It is also interesting to note that we have tackled the analysis of the two-state model by considering continuous variables, $\{p_{i}\}$, associated to each node. However, another possibility is to consider binary states for the nodes, as in Ising-like models, so that metastable states can show up  due to the multi-stable character of the node states \cite{ref1,ref2}.


\section*{Acknowledgements}
JGG acknowledges financial support from MINECO through the Ram\'on y Cajal Program Financial support came from MINECO (Projects FIS2011-25167  and FIS2012-38266-C02-01), Comunidad de Arag\'on (FENOL group). AA, JGG, MDD and SG were supported by the European Commission FET-Proactive project PLEXMATH (Grant No. 317614), and MULTIPLEX (grant 317532). AA also acknowledges financial support from the ICREA Academia and the James S.\ McDonnell Foundation, and SG and AA were supported by FIS2012-38266 and the Generalitat de Catalunya 2009-SGR-838.


\end{document}